# Generation of ground state structures and electronic properties of ternary $Al_xTi_yNi_z$ clusters ($x + y + z = 6$) with a two-stage DFT global search approach

Pin Wai Koh[1,*], Tiem Leong Yoon[1,*], Thong Leng Lim[2], Yee Hui Robin Chang[3]

and Eong Sheng Goh[1]

[1]School of Physics, Universiti Sains Malaysia, 11800 USM, Penang, Malaysia

[2]Faculty of Engineering and Technology, Multimedia University, Melaka, Malaysia

[3]Applied Sciences Faculty, Universiti Teknologi MARA, 94300 Kota Samarahan, Sarawak, Malaysia

[*]Corresponding authors

E-mail addresses: kohpinwai@gmail.com (P.W. Koh), tlyoon@usm.my (T. L. Yoon), tllim@mmu.edu.my (T. L. Lim), robincyh@sarawak.uitm.edu.my (Y. H. R. Chang), sheng5931@hotmail.com (E. S. Goh)

## Abstract

The structural and electronic properties of ternary $Al_xTi_yNi_z$ clusters, where $x$, $y$, and $z$ are integers and $x + y + z = 6$ are investigated. Both SVWN and B3LYP exchange-correlation functionals are employed in a two-stage density functional theory (DFT) calculations to generate these clusters. In the first stage, a minimum energy cluster structure is generated by an unbiased global search algorithm coupled with a DFT code using a light exchange-correlation functional and small basis sets. In the second stage, the obtained cluster structure is further optimized by another round of global minimization search coupled with a DFT calculator using a heavier exchange-correlation functional and more costly basis set. Electronic properties of the structures are illustrated in the form of a ternary diagram. Our DFT calculations find that the <mark>thermodynamic stability</mark> of the clusters increases with the increment in the number of constituent nickel atoms. These results provide a new insight to the structure, stability, chemical order and electronic properties for the ternary alloy nanoclusters.

## I. Introduction

Atomic clusters are aggregates of atoms ranging from a few to thousands of atoms or molecules. Nanoclusters are atomic clusters with a diameter in the order of nanometers. They exhibit distinctly different electronic and structural behaviors compared to their larger size counterpart due to low dimensional and quantum confinement effects [1]. From the year 2000 onwards, transition metal clusters had been intensively studied, both experimentally [2]–[7] and computationally [8]–[31]. Nanoclusters, mainly binaries or ternaries, have attracted much attention due to their broad applications in catalysis [32]–[34], magnetic-recording materials [35] and biological applications, to name a few. For example, FeAlAu$_n$ ($n = 1-6$) [36], Fe-Co-Ni [1],[39],[40], Fe-Co-Pd [39] and Ag-Au-Pd [40] trimetallic clusters have been studied for their magnetic, electronic, and structural properties.

In the search of the ground state structures of ternary alloy clusters, one common practice is to generate them based on classical and semi-classical methods such as adoption of Gupta potential, Sutton-Chen potential, and others empirical potentials. These empirical or semi-empirical results commonly show that the ground state structures of the small clusters are in the shape of an icosahedron, whereas truncated octahedron and a truncated decahedral structure is favored by the large clusters [2]. Structural evolution of cluster can be explained and tackled by classical and semi-classical approaches but these methods may fail if the electronic effects from valence electrons of the atoms have to be taken into account [1], [39], [40]. Using classical and semiclassical approaches in the search of ground state configurations for transition metal clusters will produce unreliable results, due to the existence of localized d orbitals [41]–[44].

Electronic structure and stability of transition metal clusters, such as intermediate size 3d/4d element clusters (especially 13-atoms cluster), have been studied extensively by DFT methods in the last two decades [17]–[31]. However, the simulation results fluctuate with different DFT software and optimization methods employed [41]–[44]. In DFT calculations, structural and energy values for a nanocluster might be different due to various types of exchange-correlation (XC) functional and basis set employed in the calculation, for example, $Ag_{13}$ and $Cu_{13}$ nanocluster that had been reported by applying either Gaussian orbital or plane-wave based DFT [25]–[29]. DFT results also vary with inclusion or non-inclusion of semi-core states in the pseudopotential [35], [45].

Currently, a few theoretical works on the binary alloy Al-Ti, Al-Ni and Ti-Ni small size nanoclusters can be found. Based on theoretical and experimental studies on the Al-Ti, Al-Ni and Ni-Ti binary alloy systems [46], [47], [48], [49] [50], [51] and ternary alloy system Al-Ti-Ni [52]–[54] , binary and ternary alloy clusters might have potential to act as potential catalyst in industrial engineering [44]. Researches done using DFT includes: aluminum-doped titanium cluster $AlTi_n$ ($n = 1 - 13$) by Xiang et al. [55], titanium-doped aluminum cluster $Al_nTi$ ($n = 2 - 24$) by Hua et al. [41], electronic and structural properties of Al-Ni cluster ($n < 5$) by Zhao et al. [43] and bimetallic Ti-Ni clusters ($n < 13$) by Chen et al. [42]. In contrast, literature about global search and generation of ground state structures of trimetallic clusters by employing full ab-initio method is very scarce. Trimetallic nanoclusters $Fe_xCo_yPd_z$ ($x + y + z = 7$) [39] and $Fe_xCo_yNi_z$ ($x + y + z = 5, 6, 7, 13$) [1], [37], [38] were studied for its interesting electronic and magnetic properties.

The structural and electronic properties of $Al_kTi_lNi_m$ ($k + l + m = 2, 3, 4$) [44], [56] and $Al_nTi_nNi_n$ ($n = 1 - 16$) [57] clusters had been investigated by Erkoc and Oymak [56]. Al-Ti-Ni cluster structures are generated by these authors based on a molecular dynamics (MD) scheme that applies Lennard-Jones (for two body part) and Axiltod-Teller triple-dipole potentials (for the three-body part) [58], whereas the electronic properties of the obtained structures are evaluated via DFT calculations within the Becke three-parameter Lee-Yang-Parr (B3LYP) and effective core potential level.

Complementing the work done by these authors, an unbiased search for the ground states structures of $Al_kTi_lNi_m$ clusters employing full DFT calculations has been carried out by the present authors recently (Koh et. al)[59]. The present paper is a natural continuation of our work which stops at the cluster size of 4 atoms. In our previous work, a two-stage computational strategy involving DFT calculations subjected to XC functional with a different computational cost at each stage was deployed to obtain the ground state structures for Al-Ti-Ni clusters. Two global minimum search algorithms, namely, basin-hopping (BH) [10] and cut-and-splice genetic operator [60] were incorporated as an integral part of the two-stage computational strategy. DFT was the only energy calculator employed in the two-stage algorithm. The two-stage algorithm has a practical advantage over other unbiased global minimum search for ground state structures of multi-elements clusters. In contrast to other global search algorithms that apply density functional tight-binding theory (DFTB) [61], [62] or MD [12], [13], [15] as energy calculator, this calculation strategy does not require Slater-Koster files and empirical potential. In order to provide a robustness check to the correctness and accuracy of the $Al_kTi_lNi_m$ clusters generated using the two-stage global search cum DFT strategy against the published results of the same clusters in the literature, we need to go beyond the small cluster size 4 atoms, which only offers a small number of distinct atomic composition. In this paper, we wish to further strengthen the reliability of the two-stage global search strategy with DFT as proposed in our previous work by reporting the ground state structures of $Al_xTi_yNi_z$ up to $x + y + z = 6$ atoms. Ternary clusters with 6 atoms offer a far richer variation in the atomic composition than ternary clusters with only 4 atoms. We shall also report the geometric, chemical order and electronic properties of the 6-atom Al-Ti-Ni clusters of different stoichiometries in the form of ternary diagrams.

## II. Computational Method

The capability to find the global minimum in the potential energy surface (PES) is strongly affected by the initial configuration in a global search algorithm, in which it is highly possible that iterations from an initial configuration tend to be trapped in a local minimum with a high energy barrier. Due to this reason, it is advisable that the search algorithm is initiated with a series of different initial configurations.

The computational strategy used in the present paper is based on a search algorithm that integrates basin hopping (BH) with genetic algorithm, known as the Parallel Tempering Multicanonical Basin-Hopping plus Genetic Algorithm (PTMBHGA), first proposed by Hsu and Lai [61]–[65]. Basin Hopping technique is an unbiased optimization method that is introduced by Wales and Doye [10] and Li and Scheraga [11]. This optimization approach has been widely employed in numerous theoretical works to locate the ground state structure or a global minimum energy state of an atomic cluster system. BH makes use a genetic-like operation known as angular move or random displacement (AMRD) [9], [10], [61]–[64]. AMRD is a random move method to alter the positions of the cluster structure and thus give birth to a new configuration. An advantage of using this method is it effectively helps to discover the optimized energy value for potential energy functions with a funnel landscape.

Apart from BH, the PTMBHGA code also includes another powerful unbiased search algorithm, the well-known genetic algorithm (GA). GA starts with a population of initial (guesses) candidates, known as "parents". A selection process is stipulated and applied to sort out the best candidates among the parents and discard the remaining ones based on the fitness [63], [64] of each candidate. A genetic operator is then invoked to generate new individuals (children) as subsequent replacement of the discarded parents. The process is repeated until the best collection of individuals is found and the global energy minimum is presumably contained in this collection. The version of PTMBHGA code we use for this work provides a selection of 7 genetic operators (GOs). However, after some initial trial-and-error calculations it is realized that the only essential operator effective for the present system (which is a multiple component alloy cluster) is the cut and splice GO. Its operation employs and cut and splice technique to generate new structure configurations from a previous one [60].

Our previous work [59], as well as the present one, use a modified version of the original PTMBHGA code for generating and optimizing the configurations of Al-Ti-Ni multi-component alloy clusters. The PTMBHGA code used in this work is interfaced with the first-principles DFT package Gaussian 09 (G09) [65]. The modified PTMBHGA code shall be dubbed 'PTMBHGA-G09'. In the nomenclature for the unbiased search of ground state structures of clusters, PTMBHGA is referred as the 'global optimizer', while G09 the 'energy calculator' [66].

Recall that our main objective is to obtain the lowest energy configuration of an $Al_xTi_yNi_z$ cluster ($x + y + z = 6$) with arbitrary composition at the DFT level, initiated from a random initial configuration. To this end, the calculation procedure is divided into two stages. In the first stage, low-lying structures (LLS) are generated to populate the DFT potential energy surface (PES) by using two algorithms, namely, BH and GA. The procedure proceeds as the following. Initially, 20 parent configurations are randomly generated using BH or GA at a 50:50 probability. Each of these parent configurations is locally relaxed by the default optimization algorithm in G09, which provides the total energy of the locally optimized configuration. This configuration will be used as an initial configuration to produce offspring configurations by using either BH (AMRD operation) or GA (cut and splice operation) at a 50:50 probability. The resultant configurations are again subjected to a local energy relaxation by G09 that determines their total energy. The DFT calculation in G09 is performed by using the Slater, Vosko, Wilks, and Nusair (SVWN) exchange-correlation functional and 3-21G Pople basis set [65]. In G09, SVWN is an XC functional equivalent to the local spin density approximation (LSDA). For the sake of convenience, we shall dub the 200-step procedure described above as the 'BH-GA generating procedure', which can be summarized as follows:

initial configuration → GA/BH → offspring configurations → locally relaxed/optimized by G09 → choose the configuration with lowest energy as the next initial configuration if there are more than one configuration → initial configuration → GA/BH → …

The process above is repeated for 200 cycles for each parent configuration. The completion of these 200 cycles for each parent configuration is termed a generation. At the end of the 200 cycles, the latter configuration is relaxed to its local minimum by the Limited-memory Broyden–Fletcher–Goldfarb–Shanno (L–BFGS) [67] and SVWN/3–21G from G09 is used in this local minimization process [62]. Within one generation, many configurations of cluster would have been generated and scanned, with their total energy calculated at the DFT level. A fixed number of configurations with the lowest energy, known as low-lying structures (LLS), are kept in a 'configuration bank' during the 200-cycle generation. 20 LLS are selected from the 'configuration bank' based on a criterial known as 'fitness', which is essentially an objective function originally defined and coded in the original PTMBHGA algorithm [63], [64]. 5 configurations with the poorest fitness score will be discarded. The deficit in the number of parent configurations will be replenished by new ones randomly generated by GA or BH. These 15 + 5 configurations are then subjected to another round of generation process as described above for 50 generations. The completion of the 50-cycle generation process marks the end of our first stage calculation procedure. At this point, the 'configuration bank' would have accumulated sufficient (hopefully) LLS that represents a reasonable sampling of the potential energy surface of the cluster at the DFT level.

In the second stage, the cluster configuration with the lowest energy from the first stage is chosen (from among the many LLS) and fed into the PTMBHGA-G09 as the only 'parent' configuration (recall that, as a case of comparison, in the first stage there are 20 parents). A 50-step 'BH-GA generating procedure' is carried out. However, in the second stage, the 'BH-GA generating procedure' uses a more expensive G09 setting as compared to that used in the first stage. Specifically, the Becke three-parameter Lee-Yang-Parr (B3LYP) exchange-correlation functional and 6–311G* basis set are deployed. In addition, the Berny optimization procedures for geometry optimization [68] in the G09 is fixed to 100 cycles (which is larger than the default value used in the first stage). Calculation of vibrational frequency using ultrafine grid, which costs additional computation resources, is concurrently carried out during each BH-GA step in the second stage. The resultant configuration at the end of the second stage is presumably the ground state structure we are seeking. The vibrational frequency calculation implemented in the second stage ensures that the final ground state structure obtained is located at a minimum instead of being in a transition state. For the ionic part, we applied the L–BFGS [67] method to perform without any constraints the optimization of the cluster's geometry [62]. There are five convergence parameters in G09, namely, maximum force, root means square force, maximum displacement, root means square displacement and threshold energy. Default values of these convergence parameters are used in all of our calculations, namely, maximum and root means square force are set to the $4.5 \times 10^{-4} \, N$ and $3 \times 10^{-4} \, N$, maximum and root means square displacement are set to $1.8 \times 10^{-3}$ Å and $1.2 \times 10^{-3}$ Å respectively. The threshold energy is set to $1 \times 10^{-9}$ Hartree.

In this paper, magnetic properties of the Al–Ti–Ni clusters are not presented. Hence, spin multiplicity for a cluster with even number of electrons is set to one (singlet), whereas the spin multiplicity is set to two (doublet) for a cluster with an odd number of electrons. According to paper [69], finding out spin multiplicities of the clusters indicates the importance effects of the clusters, but such calculations requires high computational costs.

In the present work, the size of the Al-Ti-Ni clusters is larger than that in [59]. If we were to use exactly the same simulation parameters as in the case for the 4-atom clusters, the computation time would become unbearably lengthy (subjected to the constraint of our computational resource). To render the 6-atom calculation be completed within a practically affordable time scale, the number of BH-GA steps are increased in the first stage so that a larger number of optimized structures, hence a larger area in the potential energy surface, are scanned. To compensate for the increase in the resource allocated to the first stage, the number of BH-GA steps used in the second stage is reduced. Despite the reduction of the BH-GA steps in the second stage, the calculation procedure for obtaining the ground state structure of the 6-atom clusters still costs approximately three times as lengthy as compared to that of the 4-atom clusters reported in our previous work. We found that by following the above mentioned modified tweak to the two-stage calculation strategy, the results obtained for the 6-atom clusters matched well with other published works (to be discussed in the following sections).

## III. Results and Discussion

### A. Structure: geometry and average interatomic distance (AID)

All the Al-Ti-Ni six atoms clusters are shown in Fig. 1, and their type of structures are listed in Table 1. For ternary metallic clusters, ground state structure for $Al_2Ti_2Ni_2$ generated by the two-stage algorithm agrees with the work by Erkoc and Oymak [57]. For single element clusters, $Al_6$ possesses a geometry of regular octahedron that agrees well with results published by Jones [70]. Likewise, $Ti_6$ cluster is found to have a regular octahedron shape, which agrees with the finding of Medina et al. [71], but not in agreement with the result of bicapped tetrahedron structure by Xiang et al. [55] using LSDA exchange-correlation. $Ni_6$ cluster with the lowest total energy, acquired by using the two-stage DFT method, possesses a four fused triangle geometry with an energy 0.132eV lower than the work of Parks et al. [72] and Ramirez et al. [1] who claim that isomer for global minimum energy $Ni_6$ depicts an octahedron structure. For bimetallic clusters, our simulation for $Al_5Ti$ cluster shows the similar result as Hua et al. [41] which employs tight binding genetic algorithm (TBGA) combined with DFT. However, the generated ground state structure for AlTi$_5$ cluster is in the shape of bicapped tetrahedron versus regular octahedron shape found by Xiang et al. [55] using LSDA. For Ti-Ni system, our work delivers the same result as those carried out by Chen et al. [42] for Ti$_2$Ni$_4$, Ti$_3$Ni$_3$ and Ti$_4$Ni$_2$ cluster; the TiNi$_5$ and Ti$_5$Ni clusters structures are not reported in their work. To the best of our knowledge, experimental data for small binary and ternary clusters consisting of the combination of Al, Ti, Ni atoms is unavailable, and hence unable to be compared with this work.

In Fig. 2, the average interatomic distance (AID) of the cluster as a function of atomic composition is illustrated and the AID values are presented in Table 2 (see the Supplementary Information Material) for each cluster. Ni$_6$ cluster possesses the smallest AID whereas Al$_6$ has the largest AID among all generated clusters in the Al$_x$Ti$_y$Ni$_z$ system. Among binary clusters, Al-Ti clusters have larger values of AID, especially the Al$_4$Ti$_2$ cluster which is the candidate with the second largest AID in the system. In the Al-Ni and Ti-Ni binary clusters, the AID of these clusters decreases as the number of nickel atom inside the system increases. The AID of TiNi$_5$ is the lowest among the binary clusters and is the second shortest interatomic distance in the system. Among the ternary clusters, the AID of Al$_3$Ti$_2$Ni is the largest whereas AlTiNi$_4$ has the shortest AID.

## B. Stability: binding energy per atom ($E_b$), excess energy ($E_{exc}$) and the second difference energy ($\Delta_E$)

Binding energy per atom ($E_b$) is employed to measure the total thermodynamic stability of a cluster [73]–[76]. $E_b$ of a cluster is calculated by using the equation:

$$E_b\big(\text{Al}_x\text{Ti}_y\text{Ni}_z\big) = \frac{E\big(\text{Al}_x\text{Ti}_y\text{Ni}_z\big) - xE(\text{Al}) - yE(\text{Ti}) - zE(\text{Ni})}{x + y + z}. \quad (2)$$

A cluster would be considered more stable when its $E_b$ is more negative [1], [77]. In Fig. 3, the binding energy, $E_b$ of the $\text{Al}_x\text{Ti}_y\text{Ni}_z$ clusters as a function of the atomic composition is presented by a simple ternary diagram. Values of binding energy are found to be larger for binary Al-Ni and Ti-Ni regions, especially when Ni composition is increased, or Al or Ti composition is reduced. The $\text{Al}_6$ cluster has the smallest binding energy among the entire cluster population. The binding energy of the $\text{Al}_x\text{Ti}_y\text{Ni}_z$ cluster is governed by the number of Ni atoms in the system. Apparently, binary and ternary clusters with higher Ni concentration such as $\text{Ti}_2\text{Ni}_4$, $\text{Ti}_3\text{Ni}_3$, $\text{AlTiNi}_4$, and $\text{AlTi}_2\text{Ni}_3$ clusters would display larger $E_b$, i.e., 3.28, 3.28, 3.27 and 3.25 eV respectively.

Compared to the binding energy ($E_b$), excess energy ($E_{\text{exc}}$) is a parameter more sensible to the geometries of clusters. Second order difference energy ($\Delta_E$) is another crucial parameter to determine against the replacement of an element by a different one in a system. Calculations for the values of $E_b$ and $E_{\text{exc}}$ for the ternary clusters are carried out to further determine the cluster formation with the given composition, which is known as possible magic compositions. Ferrando et al. [40] and Granja et al. [1], [77] have used a similar quantitative approach to access the cases of binary and ternary clusters respectively. $E_{\text{exc}}$ and $\Delta_E$ can be calculated for the $\text{Al}_x\text{Ti}_y\text{Ni}_z$ system as follows ($x + y + z = N$):

$$E_{\text{exc}}(\text{Al}_x\text{Ti}_y\text{Ni}_z) = E_b(\text{Al}_x\text{Ti}_y\text{Ni}_z) - x\frac{E_b(\text{Al}_N)}{N} - y\frac{E_b(\text{Ti}_N)}{N} - z\frac{E_b(\text{Ni}_N)}{N}, \quad (3)$$

$$\Delta_E(\text{Al}_x\text{Ti}_y\text{Ni}_z) = \frac{1}{n_{xyz}}[E_b\big(\text{Al}_{x+1}\text{Ti}_{y-1}\text{Ni}_z\big) \\ + E_b\big(\text{Al}_{x-1}\text{Ti}_{y+1}\text{Ni}_z\big) + E_b\big(\text{Al}_{x+1}\text{Ti}_y\text{Ni}_{z-1}\big) \\ + E_b(\text{Al}_{x-1}\text{Ti}_y\text{Ni}_{z+1}) + E_b(\text{Al}_x\text{Ti}_{y+1}\text{Ni}_{z-1})$$

$$\quad (4)$$

$$+ E_b(\text{Al}_x\text{Ti}_{y-1}\text{Ni}_{z+1}) - n_{xyz}E_b(\text{Al}_x\text{Ti}_y\text{Ni}_z)],$$

where $n_{xyz}$ and $1/n_{xyz}$ is the total number of the nearest neighbor structure and normalization factor of a cluster, respectively. The normalization factor $(1/n_{xyz})$ of $\Delta_E$ is equal to two for single element clusters, four for binary clusters and six for ternary clusters to ensure a better comparison for pure, binary and ternary clusters.

The values of binding energy per atom, excessive energy, and second-order difference energy are tabled in Table 2. Clusters with more negative $E_{\text{exc}}$ value tend to be mixed while pure element clusters possess zero excess energy, $E_{\text{exc}} = 0$ (it tends to segregate); $\text{Al}_N$, $\text{Ti}_N$, and $\text{Ni}_N$ single element clusters are less preferable than the ternary and binary clusters in cluster formation. A cluster with the most negative $\Delta_E$ value also infers that it retains high relative stability and it also can be considered as a magic composition. In Figs. 4 and 5, the values obtained for $E_{\text{exc}}$ and $\Delta_E$ are plotted to display the stability of six-atom clusters (refers to Fig. 1). As the number of heteronuclear bonds within the cluster increases, its stability increases, which reflects elements in the cluster tends to be mixed rather than segregated. From Figs. 4 and 5, both $E_{\text{exc}}$ and $\Delta_E$ exhibit a similar trend with the composition. In Fig. 4 and Fig. 5, binary Al-Ni and Ti-Ni, and the Ni-rich ternary compositions tend to possess larger values of $E_{\text{exc}}$ and $\Delta_E$. This is observed in Co-Ni clusters as well [1], [37], [38], which are very near to the pure Ni and Co compositions in the ternary diagram and are the least favored candidates for alloy characterization. Although the value of $E_b$ for the AlNi$_5$ is slightly lower than the Ti$_2$Ni$_4$ cluster, values of $E_{\text{exc}}$ and $\Delta_E$ for AlNi$_5$ are much larger than the Ti$_2$Ni$_4$ cluster. Comparing the minima obtained for $E_{\text{exc}}$ and maxima for $\Delta_E$, the most favorable binary and ternary alloys clusters are AlNi$_5$ and AlTiNi$_4$. This is supported by the fact that these alloys are found abundantly in cluster growing experiment [37].

## C. Chemical Order

Segregation or mixing phenomenon (also known as chemical order) for all the $Al_xTi_yNi_z$ cluster configurations is studied to discern the mutual influence of the multi-component alloy structure. Chemical order is a parameter introduced by Ducastelle [78] to study bulk-like binary alloy systems. A clear distinction between disorder and mixing [37] is displayed by bulk-like binary alloy systems when its chemical order value, $\sigma$, approximates zero and small negative, respectively. Ordered phases such as layered-like phase may emerge in the bulk-like binary alloy systems when $\sigma$ is a large negative value [78]. Chemical order $\sigma$ as a function of the relative composition has the following characteristics: positive when homoatomic pairs dominate over the heteroatomic pairs, which means that segregation or phase separation takes place in a cluster; negative when mixing is present, indicating that hetero-atomic pairs are more prominent in the cluster. If the value of chemical order approximates zero ($\sigma \approx 0$), this implies that the cluster undergoes a phase transition from segregation to mixing or vice-versa.

Based on several literature reviews [38], [40], [78], the chemical order parameter ($\sigma$) in our case can be defined as follows:

$$\sigma = \frac{n_{Al-Al} + n_{Ti-Ti} + n_{Ni-Ni} - n_{Al-Ti} - n_{Al-Ni} - n_{Ti-Ni}}{n_{Al-Al} + n_{Ti-Ti} + n_{Ni-Ni} + n_{Al-Ti} + n_{Al-Ni} + n_{Ti-Ni}}, \qquad (5)$$

where $n_{A-B}$ is the number of nearest $A-B$ bonds (see the column of $N-N$ pairs distribution in Table 3, Supplementary Information Material).

The order parameters $\sigma$ for all the cluster configurations are given in Table 3 and Fig. 6. As expected, segregation ($\sigma = 1$) is observed near the corner of the triangle, i.e., in the region where all the pure elements clusters are located, e.g. $Al_6$, $Ti_6$, and $Ni_6$ clusters. Bimetallic clusters $Al_5Ni$, $TiNi_5$, and $Ti_4Ni_2$ clusters display zero order parameter, indicating a transition between segregation and mixing. The mixing phase (with negative $\sigma$ value) are located mainly at the central region of the triangle (also known as a ternary region), inferring that ternary clusters prefer mixing and their stability increases with more heterogeneous bonds.

## D. Electronic Properties: ionization potential (IP), electron affinity (EA), Global hardness ($\eta$), Mulliken electronegativity ($\xi$), HOMO-LUMO energy gaps ($E_{gap}$) and Polarizability ($\alpha$)

Fig. 7 and Fig. 8 show the ionization potential (IP) and electron affinity (EA) for $Al_xTi_yNi_z$ clusters. Definitions of ionization potential (IP) and electron affinity (EA) in the context of the $Al_xTi_yNi_z$ clusters are given in the Supplementary Information Material. Electrons are difficult to be removed from a neutral cluster when a cluster obtains a higher IP value. A cluster with higher EA indicates that a large amount of energy is released when an electron is added to a neutral cluster. In Fig. 7, high values of IP are observed at Al-Ni edge of the ternary diagram, and $AlNi_5$ cluster acquires the highest IP among all the clusters. Highest value of IP also means that it is very hard to expel an electron from the $AlNi_5$ cluster in order to form a cationic cluster. However, IPs of all the ternary clusters are slightly lower when compared to the pure element $Ni_6$ cluster and binary clusters. The entire pure element clusters exhibit high EA values. High values of EA are displayed along the Ti-Ni edge, especially $TiNi_5$ cluster that possesses highest value of EA among all the clusters. $AlNi_5$ and $Al_3Ni_3$ clusters are the clusters which exhibit a negative value of EA. This indicates that $AlNi_5$ and $Al_3Ni_3$ clusters are highly unstable to form an anionic cluster when an electron is added to it.

The values of IP and EA obtained are then applied to calculate global hardness ($\eta$) and Mulliken electronegativity ($\xi$) for all the cluster configurations (see the Supplementary Information Material for the definitions of $\eta$ and $\xi$). Both $\eta$ and $\xi$ parameters are shown in Fig. 9 and Fig. 10. Besides IP value, AlNi$_5$ cluster also possesses the highest value of global hardness in the system. The Mulliken electronegativity is correlated to the chemical potential ($\mu$) of the system. The largest electronegativities are in the vicinity of pure Al$_6$ cluster and ternary alloy Al$_4$TiNi cluster. Electronic data that includes ionization potential, electron affinity, global hardness and Mulliken electronegativity for the Al-Ti-Ni cluster system are reported in Table 3.

The energy difference between the HOMO (highest occupied molecular orbital) and the LUMO (lowest unoccupied molecular orbital) is known as molecular orbital energy gap ($E_{\text{gap}}$). $E_{\text{gap}}$ can be used to measure the capability of an electron to transfer from an occupied orbital to an unoccupied orbital [79]–[81]. Referring to the work by Sansores et al. [82], the overall $E_{\text{gap}}$ is defined by

$$E_{\sigma\sigma'} = |E(\text{HOMO}_{\sigma'}) - E(\text{LUMO}_\sigma)|, \qquad \sigma, \sigma' = \alpha, \beta$$
$$E_{\text{HOMO-LUMO}} = \min\{E_{\sigma\sigma'}\}. \tag{6}$$

Spin value for HOMO$_\alpha$, LUMO$_\alpha$, HOMO$_\beta$, and LUMO$_\beta$ are performed for the opened shell systems whereas spin value HOMO$_\alpha$ and LUMO$_\alpha$ are calculated for the closed shell systems. Chemical reactivity of a cluster is weak when a cluster possesses a high value of $E_{\text{gap}}$. In Fig. 11, $E_{\text{gap}}$ for all the Al$_x$Ti$_y$Ni$_z$ clusters are illustrated in the form of a ternary diagram. The bimetallic AlNi$_5$ cluster and trimetallic AlTiNi$_4$ exhibit the largest value of $E_{\text{gap}}$ along the Al–Ni and Ti–Ni edges. Larger values of the $E_{\text{gap}}$ are observed for most of the ternary clusters, which reaffirm that the clusters are stable with respect to alloying. The value of $E_{\text{gap}}$ for each cluster can also be found in Table 2. One of the most important observables for the understanding of the electronic properties of clusters is static polarizability. The mean polarizability is calculated from polarizability tensor components as:

$$\langle \alpha \rangle = \frac{1}{3}\left(\alpha_{xx} + \alpha_{yy} + \alpha_{zz}\right). \qquad (7)$$

The mean polarizability is proportional to the number of electrons of the system, and it is very sensitive to the delocalization of valence electrons as well as to the structure and shape of the system [83]. The average B3LYP/6-311G* values of static mean polarizability calculated using equation at above and mean polarizability per atom are shown in Table 4. Unfortunately, there are no measurements of static electric polarizabilities that can be compared to our system. For mono–element cluster, $Ni_6$ cluster possesses the lowest value of mean polarizability per atom, followed binary element cluster that are $TiNi_5$. Binary element such as $Ti_2Ni_4$, $AlNi_5$, $Al_2Ni_4$ clusters also have the lower mean polarizability per atom values when compared to other binary alloy clusters. The ternary alloy cluster that possesses a lowest value of mean polarizability peratom is $AlTiNi_4$. It is found that the value of mean polarizability per atom is decreased when the numbers of nickel atoms inside the Al-Ti-Ni system are increased. In Table 2 and Table 3, the unit for average interatomic distances is Angstrom (Å). The unit for binding energy per atom ($E_b$), excess energy ($E_{exc}$), second order difference energy ($\Delta_E$), HOMO–LUMO energy gaps ($E_{gap}$), ionization potential (IP), electron affinity (EA), global hardness ($\eta$) and Mulliken electronegativity ($\xi$) is electronvolt (eV). Chemical order, $\sigma$, is unitless. In Table 4, the unit of total energy ($E_T$) and summation of total energy and zero point energy ($E_T + ZPE$) is electronvolt (eV), whereas static mean polarizability ($\langle \alpha \rangle$) and mean polarizability ($\langle \alpha \rangle / N$) is atomic unit ($a.u.$). For your information, tables 2, 3 and 4 are located in Supplementary material Information.

**IV. Conclusion**

We have calculated the stability, geometric and electronic properties of ternary $Al_xTi_yNi_z$ ($x + y + z = 6$) clusters by using a two-stage procedure. In the first stage of the procedure, the basin-hopping genetic algorithm is coupled with a density functional theory calculator using a relatively cheaper basis set SVWN/3–21G, while in the second stage, the more expensive B3LYP/6–311G* basis set is used. Clusters with a high concentration of Al exhibit higher interatomic distance while those with a high concentration of Ni display smaller interatomic distance. Ni-rich clusters not only display smaller values of binding energy and excess energy, but they also exhibit larger values of second difference energy and HOMO-LUMO energy gap. Among all the clusters, AlNi$_5$ cluster possesses a maximum value of ionization potential and global hardness. Analysis based on chemical order parameter indicates that ternary $Al_xTi_yNi_z$ clusters favor mixing rather than segregation. The investigation on the structural and electronic properties of the six-atom ternary alloy nanocluster serves as an natural extension to our previous paper on 4-atom $Al_xTi_yNi_z$ clusters. The richness of the structures of the 6-atom clusters, hence the spectrum of reactivity properties, span a larger breadth than that offered by the 4-atom clusters. Although some general tendencies have been derived, further theoretical and experimental investigations of these Al-Ti-Ni clusters are required. The present work of small ternary $Al_xTi_yNi_z$ clusters could stimulate further research of physical, chemical and magnetic properties of ternary clusters, especially the magnetic properties of these particular clusters that have not yet been investigated so far.

| Pure | structure | Binary | Structure | Ternary | Structure |
|------|-----------|--------|-----------|---------|-----------|
| 006 | Four fused triangle | 015 | pentagonal pyramidal | 114 | Irregular octahedron |
| 060 | Regular octahedron | 024 | regular octahedron | 123 | edge-capped trigonal bipyramidal |
| 600 | Regular octahedron | 033 | edge-capped trigonal bipyramidal | 132 | Bicapped tetrahedron |
| | | 042 | Bicapped tetrahedron | 141 | Irregular octahedron |
| | | 051 | Bicapped tetrahedron | 213 | edge-capped trigonal bipyramidal |
| | | 105 | Bicapped tetrahedron | 222 | Bicapped tetrahedron |
| | | 204 | Bicapped tetrahedron | 231 | Bicapped tetrahedron |
| | | 303 | edge-capped tetrahedron | 312 | pentagonal pyramidal |
| | | 402 | edge-capped tetrahedron | 321 | Bicapped tetrahedron |
| | | 501 | pentagonal pyramidal | 411 | edge-capped pyramidal |
| | | 150 | Bicapped tetrahedron | | |
| | | 240 | regular octahedron | | |
| | | 330 | edge-capped trigonal bipyramidal | | |
| | | 420 | Bicapped tetrahedron | | |
| | | 510 | edge-capped pyramidal | | |

Table 1. Structures of the Al-Ti-Ni cluster are showed. Numbers in the 1st, 3rd and 5th column indicate the number of Al, Ti and Ni atoms in each cluster.

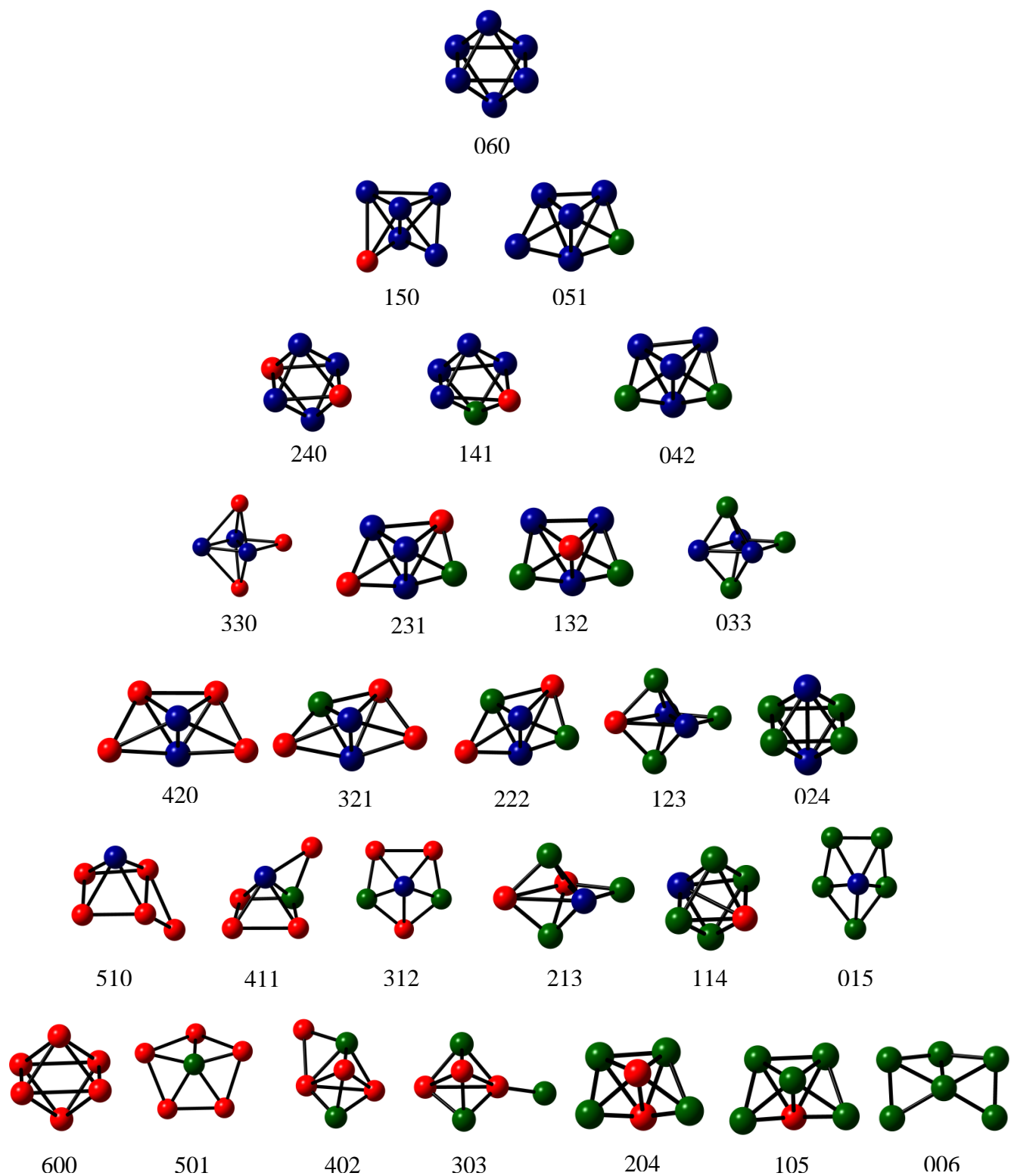

Fig. 1 Ground states structures of $Al_xTi_yNi_z$ ($x + y + z = 6$) clusters as a function of the atomic composition. Blue sphere represents Ti atoms, red for Al and green for Ni. The number below each cluster geometry model indicates the number of Ti, Ni and Al atoms in each cluster.

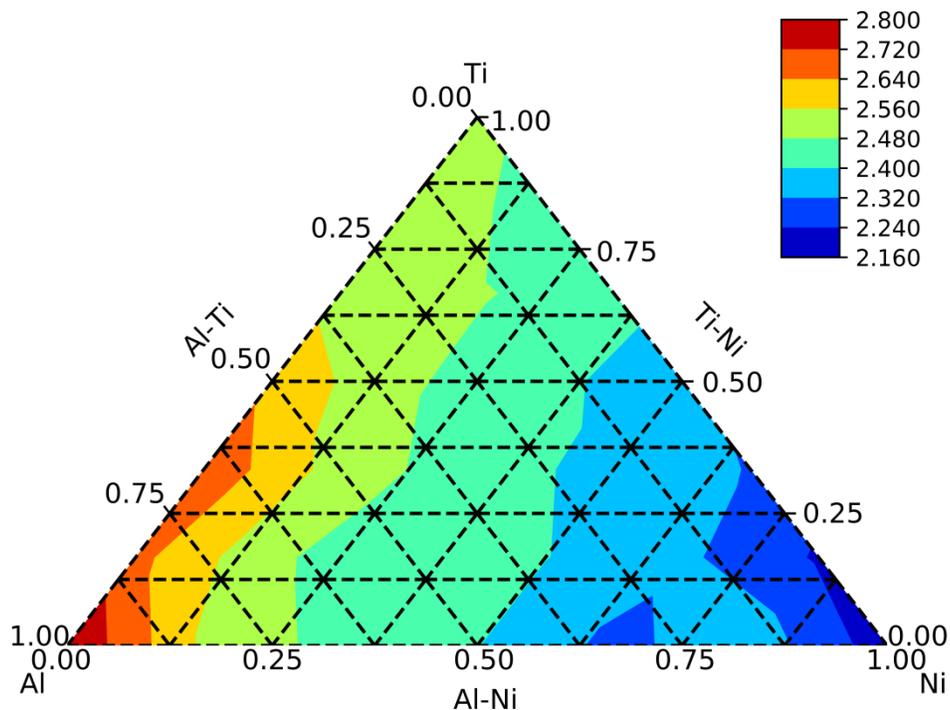

Fig. 2 Average interatomic distance of Al$_x$Ti$_y$Ni$_z$ ($x + y + z = 6$) clusters as a function of the atomic composition.

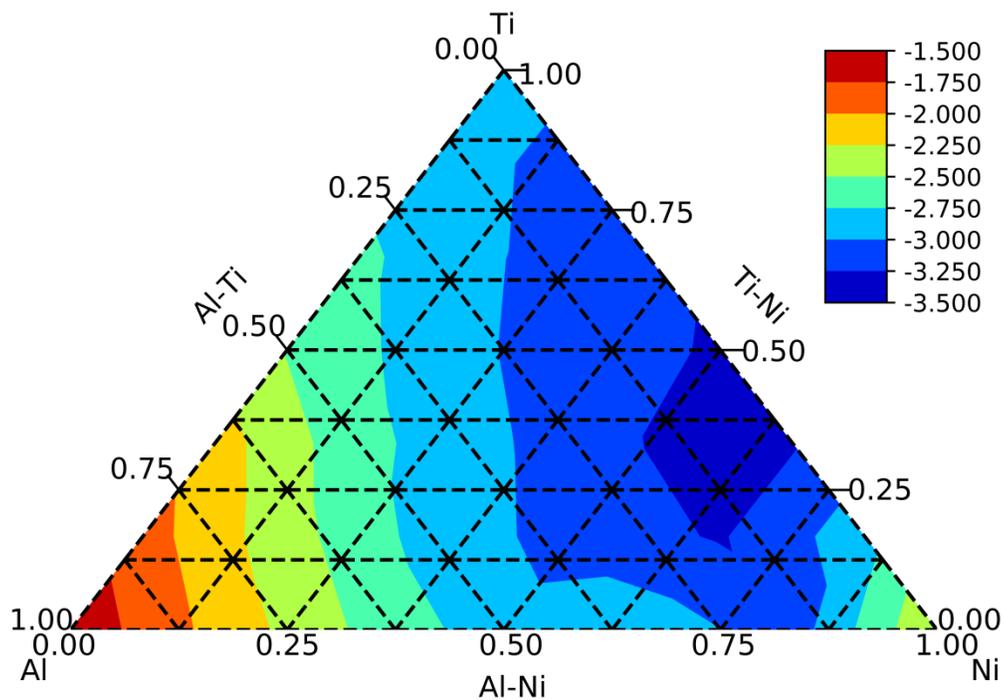

Fig. 3 Binding energy per atom of Al$_x$Ti$_y$Ni$_z$ ($x + y + z = 6$) clusters as a function of the atomic composition.

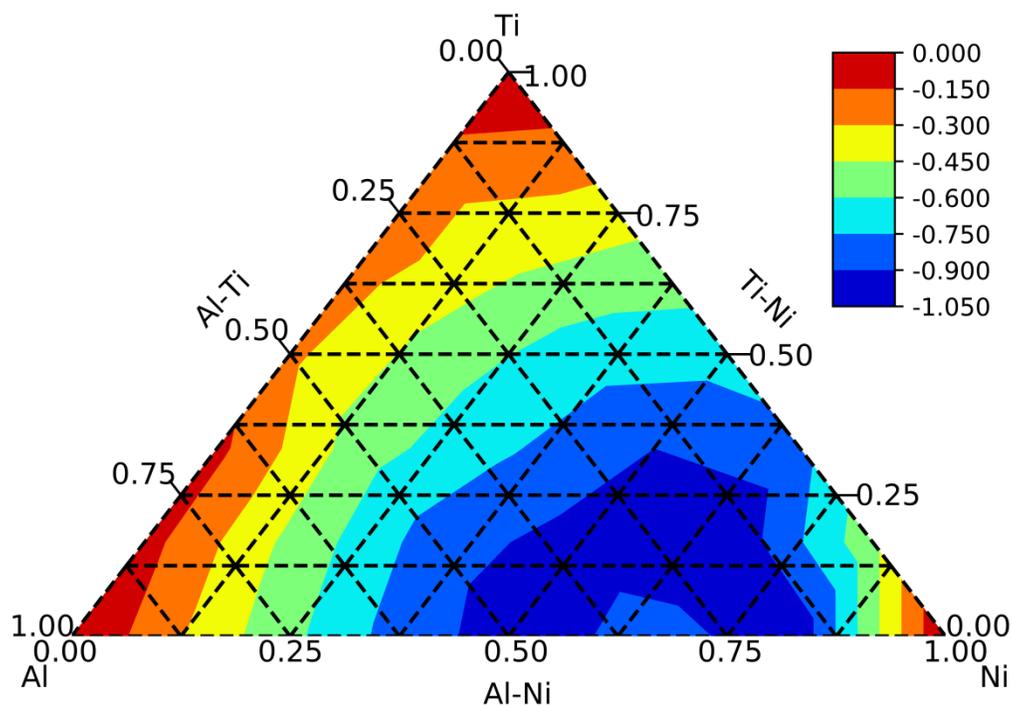

Fig. 4 Excess energy of Al$_x$Ti$_y$Ni$_z$ ($x + y + z = 6$) clusters as a function of the atomic composition.

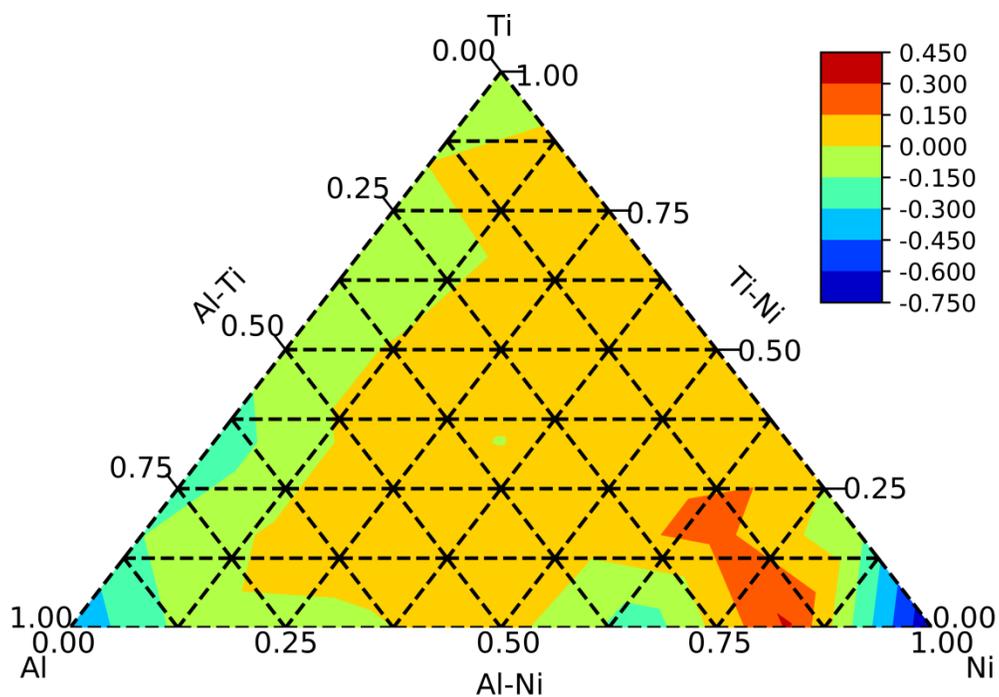

Fig. 5 Second order difference energy of Al$_x$Ti$_y$Ni$_z$ ($x + y + z = 6$) clusters as a function of the atomic composition.

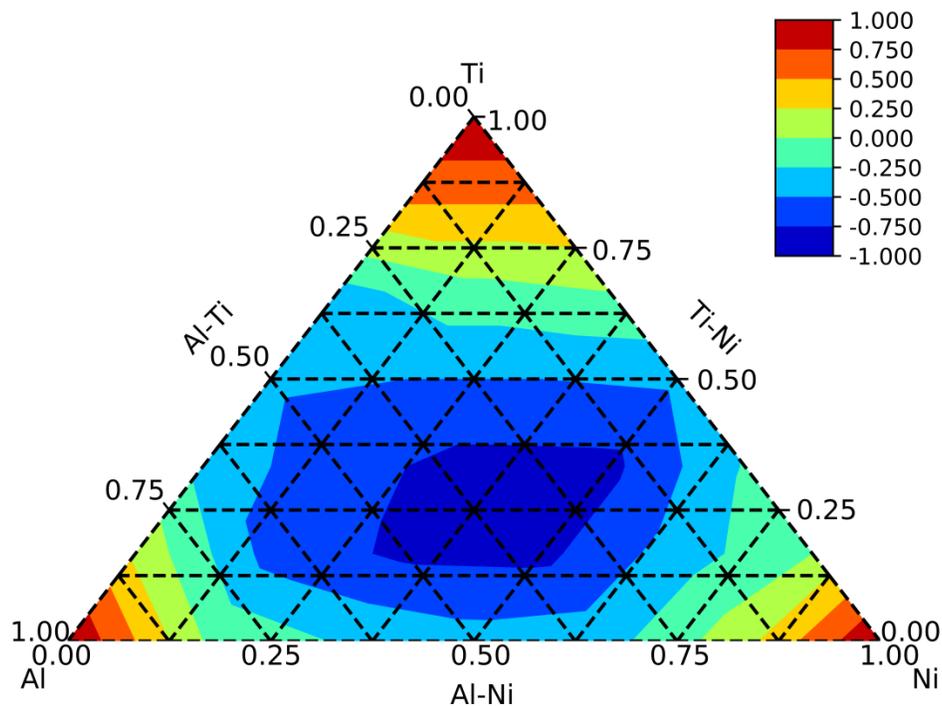

Fig. 6 Chemical order parameter ($\sigma$) of $Al_xTi_yNi_z$ ($x + y + z = 6$) clusters as a function of the atomic composition.

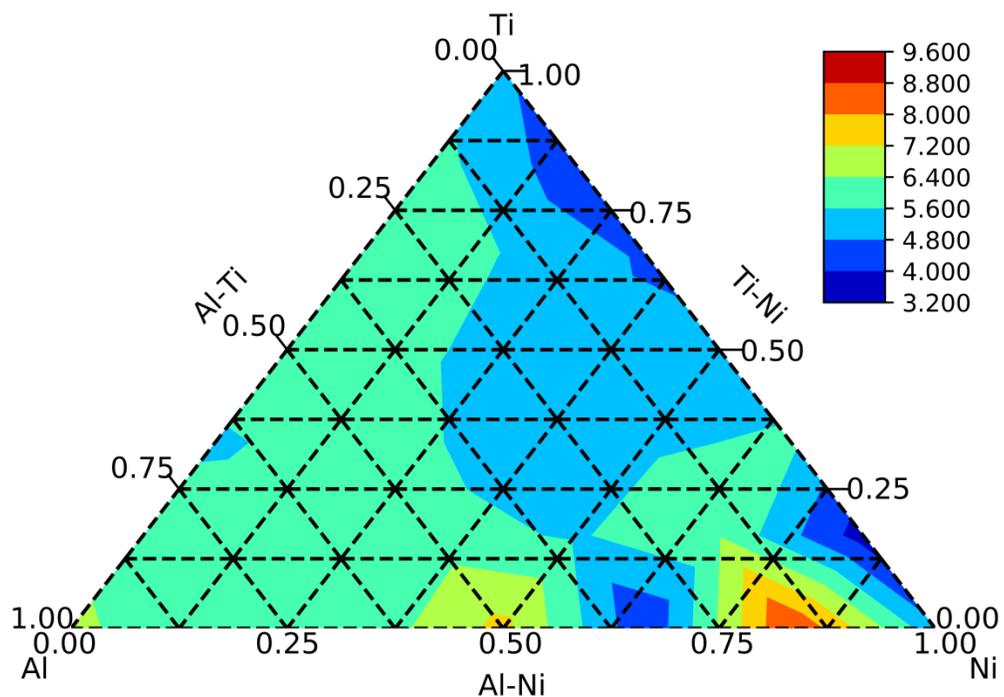

Fig. 7 Ionisation potential (IP) of $Al_xTi_yNi_z$ ($x + y + z = 6$) clusters as a function of the atomic composition.

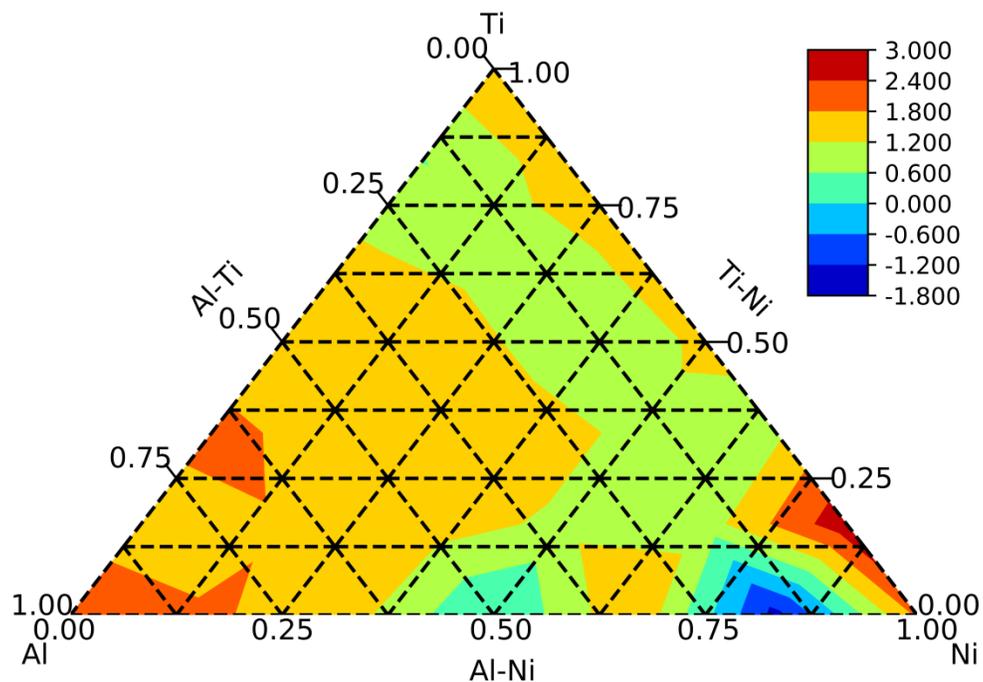

Fig. 8 Electron affinity (EA) of Al$_x$Ti$_y$Ni$_z$ ($x + y + z = 6$) clusters as a function of the atomic composition.

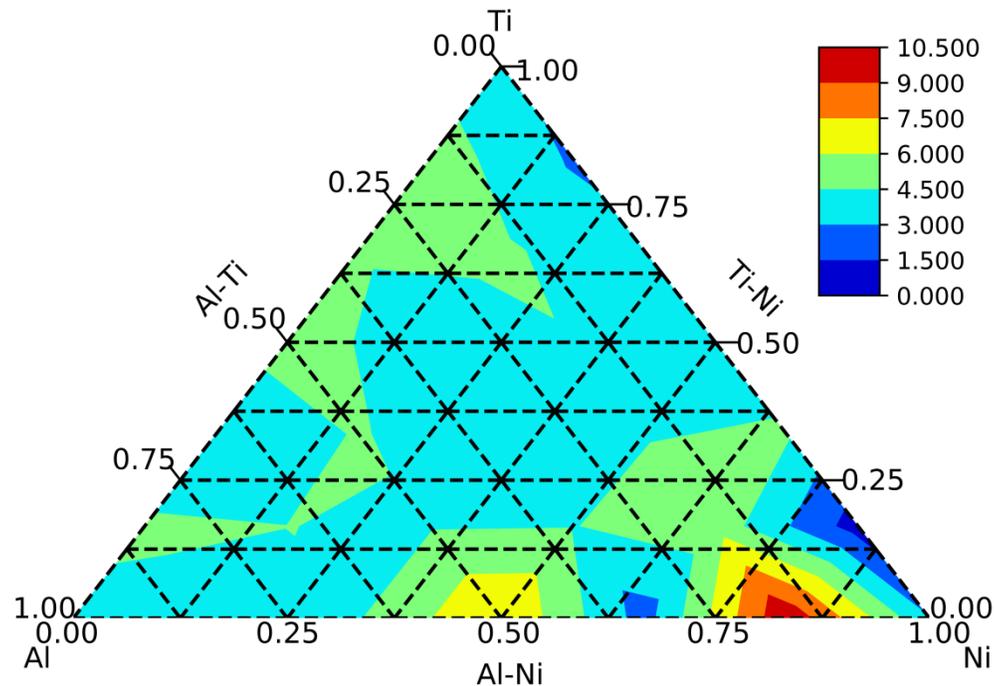

Fig. 9 Global hardness ($\eta$) of Al$_x$Ti$_y$Ni$_z$ ($x + y + z = 6$) clusters as a function of the atomic composition.

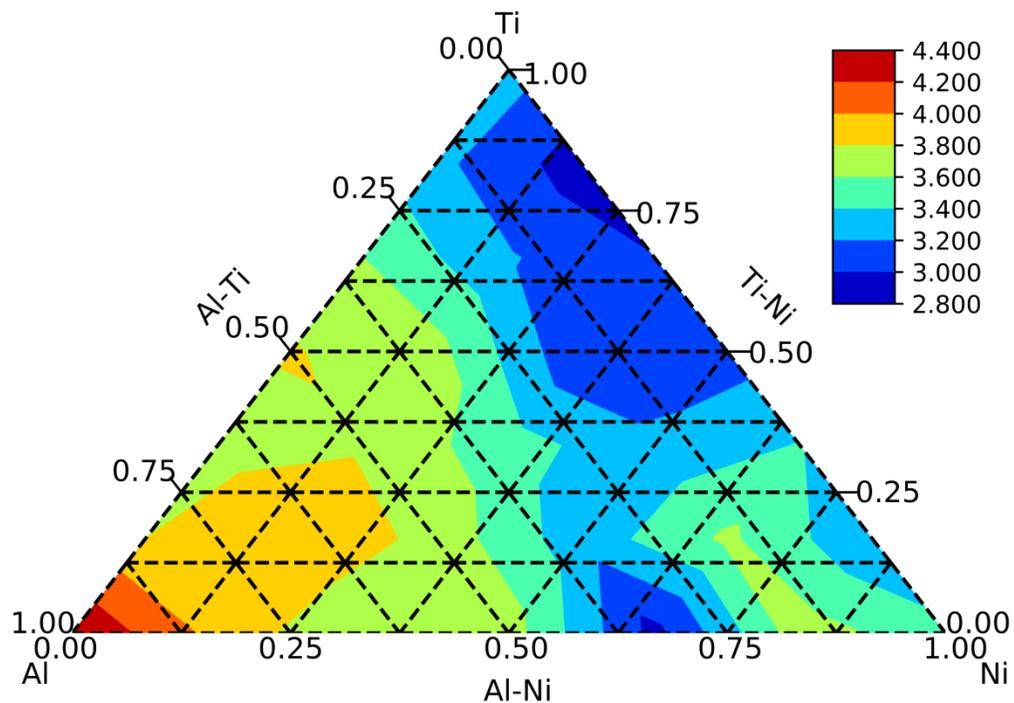

Fig. 10 Mulliken electronegativity ($\xi$) of $Al_xTi_yNi_z$ ($x + y + z = 6$) clusters as a function of the atomic composition.

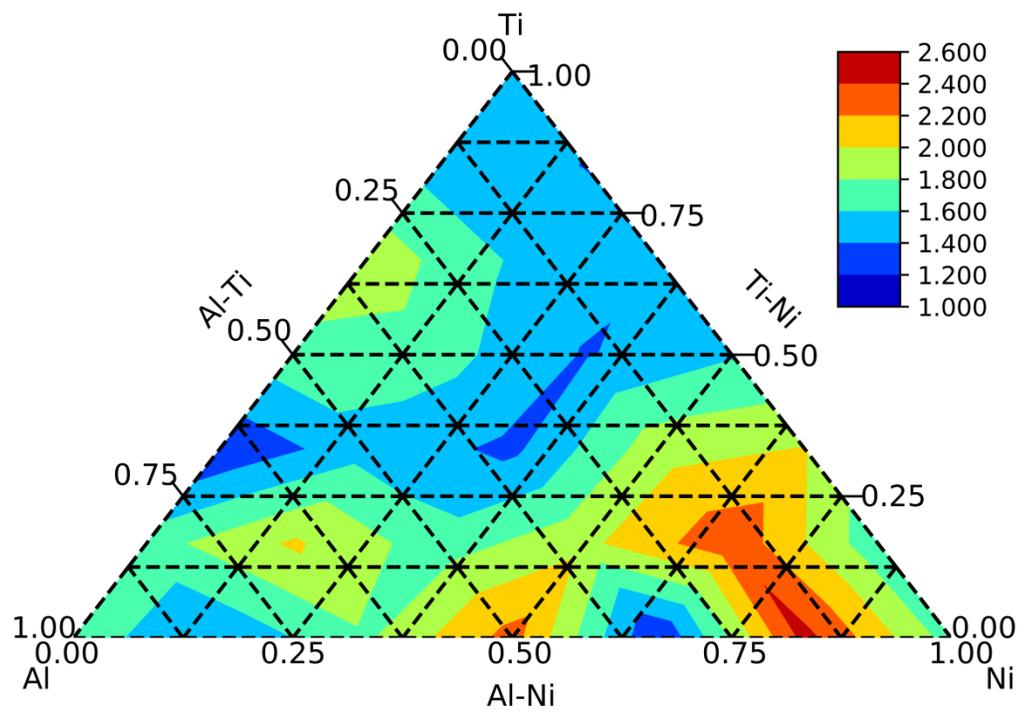

Fig. 11 HOMO-LUMO energy gaps of $Al_xTi_yNi_z$ ($x + y + z = 6$) clusters as a function of the atomic composition.

**Acknowledgments**

KPW acknowledges the financial support from the MyPhD scheme by the Malaysian Ministry of Higher Education (Kementerian Pendidikan Tinggi Malaysia). We acknowledge Prof. S. K. Lai from NCU (National Central University, Taiwan) for his tremendous support, crucial advice and provision of computational resources (including the PTMBHGA code). YTL and KPW acknowledge the financial support of USM Bridging Fund (2018).

**References**

[1] G. G. Ramírez, J. Robles, A. Vega and F. A. Granja, *J. Chem. Phys.*, vol. 134, no. 5, p. 054101, 2011.

[2] F. Baletto and R. Ferrando, *Rev. Mod. Phys.*, vol. 77, no. 1, pp. 371–423, 2005.

[3] T. P. Martin, *Phys. Rep.*, vol. 273, no. 4, pp. 199–241, 1996.

[4] E. K. Parks, L. Zhu, J. Ho, and S. J. Riley, *J. Chem. Phys.*, vol. 102, no. 19, p. 7377, 1995.

[5] E. K. Parks, G. C. Nieman, K. P. Kerns, and S. J. Riley, *J. Chem. Phys.*, vol. 107, no. 6, pp. 1861–1871, 1997.

[6] E. K. Parks, G. C. Nieman, K. P. Kerns, and S. J. Riley, *J. Chem. Phys.*, vol. 108, no. 9, pp. 3731–3739, 1998.

[7] E. K. Parks, K. P. Kerns, and S. J. Riley, *Chem. Phys.*, vol. 262, no. 1, pp. 151–167, 2000.

[8] J. A. Elliott, Y. Shibuta, and D. J. Wales, *Philos. Mag.*, vol. 89, no. 34–36, pp. 3311–3322, 2009.

[9] J. P. K. Doye and D. J. Wales, *New J. Chem.*, vol. 22, no. 7, pp. 733–744, 1998.

[10] D. J. Wales and J. P. K. Doye, *J. Phys. Chem. A*, vol. 101, no. 28, pp. 5111–5116, 1997.

[11] Z. Li and H. A. Scheraga, *J. Chem. Phys.*, vol. 92, no. 9, p. 5499, 1990.

[12] N. N. Lathiotakis, A. N. Andriotis, M. Menon, and J. Connolly, *Europhys. Lett.*, vol. 29, no. 2, pp. 135–140, 1995.

[13] V. G. Grigoryan and M. Springborg, *Chem. Phys. Lett.*, vol. 375, no. 1–2, pp. 219–226, 2003.

[14] C. Luo, *New J. Phys.*, vol. 4, pp. 10.1–10.8, 2002.

[15] J. M. Montejano-Carrizales, M. P. Iñiguez, J. A. Alonso, and M. J. López, *Phys. Rev. B*, vol. 54, no. 8, pp. 5961–5969, 1996.

[16] R. P. Gupta, *Phys. Rev. B*, vol. 23, no. 12, pp. 6265–6270, 1981.

[17] C. Massobrio, A. Pasquarello, and R. Car, *Chem. Phys. Lett.*, vol. 238, no. 4–6, pp. 215–221, 1995.

[18] P. Calaminici, A. M. Köster, N. Russo, and D. R. Salahub, *J. Chem. Phys.*, vol. 105, no. 21, pp. 9546–9556, 1996.

[19] J. Oviedo and R. E. Palmer, *J. Chem. Phys.*, vol. 117, no. 21, pp. 9548–9551, 2002.

[20] C. M. Chang and M. Y. Chou, *Phys. Rev. Lett.*, vol. 93, no. 13, pp. 133401–1–133401–4, 2004.

[21] E. Aprà, R. Ferrando, and A. Fortunelli, *Phys. Rev. B*, vol. 73, no. 20, pp. 205414–1–205414–5, 2006.

[22] R. C. Longo and L. J. Gallego, *Phys. Rev. B*, vol. 74, no. 19, pp. 193409–1–193409–4, 2006.

[23] M. Yang, K. A. Jackson, C. Koehler, T. Frauenheim, and J. Jellinek, *J. Chem. Phys.*, vol. 124, no. 2, pp. 024308–1–024308–6, 2006.

[24] L. −L.Wang and D. D. Johnson, *Phys. Rev. B*, vol. 75, no. 23, pp. 235405–1–235405–10, 2007.

[25] M. Pereiro, D. Baldomir, and J. E. Arias, *Phys. Rev. A*, vol. 75, no. 6, pp. 063204–1–063204–6, 2007.

[26] C. H. Hu, C. Chizallet, H. Toulhoat, and P. Raybaud, *Phys. Rev. B*, vol. 79, no. 19, pp. 195416–1–195416–11, 2009.

[27] F. Aguilera-Granja, A. García-Fuente, and A. Vega, *Phys. Rev. B*, vol. 78, no. 13, pp. 134425–1–134425–9, 2008.

[28] J. P. Chou, H. Y. T. Chen, C. R. Hsing, C. M. Chang, C. Cheng and C. M. Wei, *Phys. Rev. B*, vol. 80, no. 13, pp. 165412–1–165412–10, 2009.

[29] M. J. Piotrowski, P. Piquini, and J. L. F. Da Silva, *Phys. Rev. B*, vol. 81, no. 15, pp. 155446–1–155446–14, 2010.

[30] J. P. Chou, C. R. Hsing, C. M. Wei, C. Cheng, and C. M. Chang, *J. Phys. Condens. Matter*, vol. 25, no. 12, pp. 125305, 1–7, 2013.

[31] S. Datta, A. K. Raychaudhuri, and T. Saha-Dasgupta, *J. Chem. Phys.*, vol. 146, no. 16, pp. 164301–1–164301–8, 2017.

[32] C. R. Henry, *Surf. Sci. Rep.*, vol. 31, no. 7–8, pp. 231–325, 1998.

[33] M. Valden, X. Lai, and D. W. Goodman, *Science*, vol. 281, no. 5383, pp. 1647–1650, 1998.

[34] S. H. Joo, S. J. Choi, I. Oh, J. Kwak, Z. Liu, O. Terasaki, and R. Ryoo, *Nature (London)*, vol. 412, no. 6843, pp. 169–172, 2001.

[35] P. Entel, M. E. Gruner, G. Rollmann, A. Hucht, S. Sahoo, A. T. Zayak, H. C. Herper, and A. Dannenberg, *Philos. Mag.*, vol. 88, no. 18–20, pp. 2725–2737, 2008.

[36] J. −F. Zhang, M. Zhang, Y. −W. Zhao, H. −Y. Zhang. L. −N. Zhao and Y. −H. Luo, *Chin. Phys. B*, vol. 24, no. 6, pp. 067101–1–067101–7, 2015.

[37] G. G. Ramírez, P. Salvador, J. Robles, A. Vega and F. A. Granja, *Theor. Chem. Acc.*, vol. 132, no. 2, pp. 1318:1–12, 2013.

[38] A. Varas, F. A. Granja, J. Rogan and M. Kiwi, *J. Magn. Magn. Mater.*, vol. 394, pp. 325–334, 2015.

[39] A. Varas, F. A. Granja, J. Rogan and M. Kiwi, *J Nanopart Res*, vol. 18, no. 252, pp. 1–12, 2016.

[40] S. Zhao, B. Zhao, X. Z. Tian, Y. L. Ren, K. Yao, J. J. Wang, J. Liu, and Y. Ren, *J. Phys. Chem. A*, vol. 121, no. 27, pp. 5226–5236, 2017.

[41] Y. Hua, Y. Liu, G. Jiang, J. Du and J. Chen, *J. Phys. Chem. A*, vol. 117, no. 12, pp. 2590–2597, 2013.

[42] Z. G. Chen, Z. Xie, Y. C. Li, Q. M. Ma and Y. Liu, *Chin. Phys. B*, vol. 19, no. 4, pp. 043102-1-043102–8, 2010.

[43] F. Hao, Y. Zhao, X. Li and F. Liu, *J. Mol. Struct.:Theochem.*, vol. 807, no. 1–3, pp. 153–158, 2007.

[44] S. Erkoc and H. Oymak, *J. Phys. Chem. B*, vol. 107, no. 44, pp. 12118–12125, 2003.

[45] M.C. Michelini, R. P. Diez and A. H. Jubert, *Comput. Mater. Sci*, vol. 31, no. 3–4, pp. 292–298, 2004.

[46] T. Hong, T. J. W. Yang, A. J. Freeman, T. Oguchi, and J. Xu, *Phys. Rev. B*, vol. 41, no. 18, pp. 12462–12467, 1990.

[47] S. Lauer, Z. Guan, H. Wolf, and Th. Wichert, *Hyperfine Interact.*, vol. 120, no. 1–8, pp. 307–312, 1999.

[48] Y. Du and N. Clavaguera, *J. Alloys and Compd*, vol. 237, no. 1–2, pp. 20–32, 1996.

[49] M. Widom and J. A. Moriarty, *Phys. Rev. B*, vol. 58, no. 14, pp. 8967–8979, 1998.

[50] A. Pasturel, C. Colinet, D. N. Manh, A. T. Paxton, and M. van Schilfgaarde, *Phys. Rev. B*, vol. 52, no. 21, pp. 15176–15190, 1995.

[51] J. Y. Rhee, B. N. Harmon, and D. W. Lynch, *Phys. Rev. B*, vol. 59, no. 3, pp. 1878–1884, 1999.

[52] D. Farkas, D. Roqueta, A. Vilette, and K. Ternes, *Modell. Simul. Mater. Sci. Eng.*, vol. 4, no. 4, pp. 359–369, 1996.

[53] B. Huneau, P. Rogl, K. Zeng, R.S.-Fetzer, M. Bohn, and J. Bauer, *Intermetallics*, vol. 7, no. 12, pp. 1337–1345, 1999.

[54] K. Zeng, R.S. Fetzer, B. Huneau, P. Rogl, and J. Bauer, *Intermetallics*, vol. 7, no. 12, pp. 1347–1359, 1999.

[55] J. Xiang, S. H. Wei, X. H. Yan, J. Q. You, and Y. L. Mao, *J. Chem. Phys.*, vol. 120, no. 9, pp. 4521–4527, 2004.

[56] H. Oymak and S. Erkoc, *Phys. ReV. A*, vol. 66, no. 3, pp. 033202-1-033202–10, 2002.

[57] H. Oymak and S. Erkoc, *Model. Simul. Mater. Sci. Eng.*, vol. 12, no. 1, pp. 109–120, 2004.

[58] B. M. Axilrod and E. J. Teller, *Chem. Phys.*, vol. 11, no. 6, pp. 299–300, 1943.

[59] P. W. Koh, T. L. Yoon, T. L. Lim and Y. H. Robin Chang, *Int. J. Quantum Chem.*, pp. 1–13, 2018.

[60] D. M. Deaven and K. M. Ho, *Phys. Rev. Lett.*, vol. 75, no. 2, pp. 288–231, 1995.

[61] T. W. Yen and S. K. Lai, *J. Chem. Phys.*, vol. 142, no. 8, pp. 084313-1-084313–13, 2015.

[62] T. W. Yen, T. L. Lim, T. L. Yoon and S. K. Lai, *Comput. Phys. Commun.*, vol. 220, pp. 143–149, 2017.

[63] S. K. Lai, P. J. Hsu, K. L. Wu, K. Liu, M. Iwamatsu, *J. Chem. Phys.*, vol. 117, no. 23, pp. 10715–10725, 2002.

[64] P. J. Hsu and S. K. Lai, *J. Chem. Phys.*, vol. 124, no. 4, pp. 044711-1-044711–11, 2006.

[65] M. J. Frisch, G. W. Trucks, H. B. Schlegel, G. E. Scuseria, M. A. Robb, J. R. Cheeseman, J. A. Montgomery, T. Vreven (Jr), K. N. Kudin, J. C. Burant, J. M. Millam, S. S. Iyengar, J, Tomasi, V. Barone, B. Mennucci, M. Cossi, G. Scalmani, N. Rega, G. A. Petersson, H. Nakatsuji, M. Hada, M. Ehara, H. Toyota,R. Fukuda, J. Hasegawa, M. Ishida, T. Nakajima, Y. Honda, O. Kitao, H. Nakai, M .Klene, X. Li, J. E. Knox, H. P. Hratchian, J. B. Cross, C. Adamo, J. Jaramillo, R. Gomperts, R. E. Stratmann, O. Yazyev, A. J. Austin, R. Cammi, C. Pomelli, J. W. Ochterski, P. Y. Ayala, K. Morokuma, G. A. Voth, P. Salvador, J. J. Dannenberg, V. G. Zakrzewski, S. Dapprich, A. D. Daniels, M. C. Strain, O. Farkas, D. K. Malick, A. D. Rabuck, K. Raghavachari, J. B. Foresman, J. V. Ortiz, Q. Cui, A. G. Baboul, S. Clifford, J. Cioslowski, B. Stefanov, B. G. Liu, A. Liashenko, P. Piskorz, I. Komaromi, R. L. Martin, D. J. Fox, T. Keith, M/ A. Al-Laham, C. Y. Peng, A. Nanayakkara, M. Challacombe, P. M. W. Gill, B. Johnson, W. and Chen, M. W. Wong, C. Gonzalez, J. A. Pople, Gaussian Inc.: Pittsburgh, PA, 2009.

[66] W. C. Ng, T. L. Lim, and T. L. Yoon, *J. Chem. Inf. Model*, vol. 57, no. 3, pp. 517–528, 2017.

[67] D. C. Liu and J. Nocedal, *Math. Program*, vol. 45, no. 1–3, pp. 503–528, 1989.

[68] H. B. Schegel, *J. Comput. Chem.*, vol. 3, no. 2, pp. 214–218, 1985.

[69] J. B. A. Davis, A. Shayeghi, S. L. Horswell and R. L. Johnston, *Nanoscale*, vol. Nanoscale, 7(33), ., no. 33, pp. 14032–14038, 2015.

[70] R. O. Jones, *Phys. Rev. Lett.*, vol. 67, no. 2, pp. 224–227, 1991.

[71] J. Medina, RdeCoss, A. Tapia, G. Canto, *Eur. Phys. J. B*, vol. 76, no. 3, pp. 427–433, 2010.

[72] E. K. Parks, L. Zhu, J. Ho, S. J. Riley, *J. Chem. Phys.*, vol. 100, no. 10, pp. 7206–7222, 1994.

[73] W. G. Sun, J. J Wang, C. Lu, X. X. Xia, X. Y. Kuang and A. Hermann, *Inorg. Chem.*, vol. 56, no. 3, pp. 1241–1248, 2017.

[74] X. D. Xing, A. Hermann, X. Y. Kuang, J. Meng, L. Cheng and Y. Y. Jin, *Sci Rep.*, vol. 6, no. 19656, pp. 1–11, 2016.

[75] X. X. Xia, A. Hermann, X. Y. Kuang, Y. Y. Jin, L. Cheng and X. D. Xing, *J. Phys. Chem. C*, vol. 120, no. 1, pp. 677–684, 2016.

[76] Y. Tian, D. Wei, Y. Y. Jin, , J. Barroso, C. Lu and G. Merino, *Phys. Chem. Chem. Phys.*, pp. 1–7, 2019.

[77] R. Ferrando, J. Jellinek and R. L. Johnston, *Chem. Rev.*, vol. 108, no. 3, pp. 845–910, 2008.

[78] F. Ducastelle, F.R. de Boer and D. G. Petifor (Eds.), *Interatomic Potential and Structural Stability*, vol. 114. Springer, Berlin, Heidelberg, 1993.

[79] Y. Y. Jin, G. Maroulis, X. Y. Kuang, L. P. Ding, C. Lu, J. J. Wang, J. Lv, C. Z. Zhang and M. Ju, *Phys Chem Chem Phys.*, vol. 17, no. 20, pp. 1–8, 2015.

[80] Y. Y. Jin, Y. H. Tian, X. Y. Kuang, C. Z. Chang, C. Lu, JJ. Wang, J. Lv and M. Ju, *J. Phys. Chem. A*, vol. 119, no. 25, pp. 6738–6745, 2015.

[81] M. Ju, J. Lv, Y. X. Kuang, P. L. Ding, C. Lu, J. J. Wang, Y. Y. Jin and G. Maroulis, *RSC Adv.*, vol. 5, no. 9, pp. 6560–6570, 2015.

[82] A. Miralrio and L. E. Sansores, *Int. J. Quantum Chem.*, vol. 114, no. 14, pp. 931–942, 2014.

[83] P. Jaque and  A. Toro-Labbé, *J. Chem. Phys.*, vol. 117, no. 7, pp. 3208–3217.

**Supplementary Information Material**

**Generation of ground state structures and electronic properties of ternary $Al_xTi_yNi_z$ clusters ($x + y + z = 6$) with a two-stage DFT global search approach**

Pin Wai Koh[1,*], Tiem Leong Yoon[1,*], Thong Leng Lim[2], Yee Hui Robin Chang[3]

and Eong Sheng Goh[1]

[1]School of Physics, Universiti Sains Malaysia, 11800 USM, Penang, Malaysia

[2]Faculty of Engineering and Technology, Multimedia University, Melaka, Malaysia

[3]Applied Sciences Faculty, Universiti Teknologi MARA, 94300 Kota Samarahan, Sarawak,

Malaysia

[*]Corresponding authors

E-mail addresses: kohpinwai@gmail.com (P.W. Koh), tlyoon@usm.my (T. L. Yoon),

tllim@mmu.edu.my (T. L. Lim), robincyh@sarawak.uitm.edu.my (Y. H. R. Chang),

sheng5931@hotmail.com (E. S. Goh)

| Alloy | Symmetry | AID | $E_b$ | $E_{\mathrm{exc}}$ | $\Delta_E$ | $E_{\mathrm{gap}}$ |
|-------|----------|-----|-------|--------|-----------|---------|
| 006 | $C_s$ | 2.1974 | −2.2485 | 0 | −0.7447 | 1.7108 |
| 060 | $D_{4h}$ | 2.5133 | −2.9185 | 0 | −0.0707 | 1.4716 |
| 600 | $D_{3d}$ | 2.7920 | −1.5734 | 0 | −0.4090 | 1.6912 |
| 015 | $C_s$ | 2.2306 | −2.8561 | −0.4959 | −0.1261 | 1.6648 |
| 024 | $D_{4h}$ | 2.3159 | −3.2757 | −0.8039 | 0.1096 | 2.0109 |
| 033 | $C_{2v}$ | 2.3333 | −3.2754 | −0.6920 | 0.0554 | 1.5772 |
| 042 | $C_{2v}$ | 2.4401 | −3.2003 | −0.5051 | 0.0816 | 1.4305 |
| 051 | $C_s$ | 2.4302 | −3.0586 | −0.2517 | 0.0500 | 1.3949 |
| 105 | $C_s$ | 2.3705 | −3.1303 | −0.9943 | 0.3200 | 2.4689 |
| 204 | $C_{2v}$ | 2.2984 | −2.8625 | −0.8391 | −0.2387 | 1.1962 |
| 303 | $C_{2v}$ | 2.4025 | −2.8973 | −0.9863 | 0.0590 | 2.2765 |
| 402 | $C_s$ | 2.4501 | −2.5357 | −0.7373 | −0.0200 | 1.7679 |
| 501 | $C_s$ | 2.5454 | −2.0735 | −0.3876 | −0.0268 | 1.4322 |
| 150 | $C_s$ | 2.5616 | −2.9198 | −0.2255 | 0.0020 | 1.5100 |
| 240 | $D_{4h}$ | 2.5313 | −2.6985 | −0.2284 | −0.1280 | 1.9369 |
| 330 | $C_s$ | 2.6195 | −2.5140 | −0.2680 | −0.0810 | 1.7266 |
| 420 | $C_{2v}$ | 2.6999 | −2.1343 | −0.1126 | −0.2348 | 1.2308 |
| 510 | $C_s$ | 2.6634 | −1.8912 | −0.0936 | −0.1543 | 1.7140 |
| 114 | $C_s$ | 2.3364 | −3.2740 | −1.0264 | 0.1929 | 2.3459 |
| 123 | $C_s$ | 2.3755 | −3.2590 | −0.8996 | 0.0831 | 1.9130 |
| 132 | $C_s$ | 2.4276 | −3.1450 | −0.6741 | 0.0470 | 1.3837 |
| 141 | $C_s$ | 2.4871 | −2.9957 | −0.4130 | 0.0125 | 1.5780 |
| 213 | $C_s$ | 2.4023 | −3.1035 | −0.9684 | 0.0827 | 1.9516 |
| 222 | $C_s$ | 2.4331 | −2.9811 | −0.7343 | −0.0032 | 1.3557 |
| 231 | $C_s$ | 2.4892 | −2.8767 | −0.5182 | 0.0426 | 1.6765 |
| 312 | $C_s$ | 2.4118 | −2.8511 | −0.8285 | 0.0863 | 1.6923 |
| 321 | $C_s$ | 2.5242 | −2.6705 | −0.5362 | 0.0442 | 1.5200 |
| 411 | $C_s$ | 2.4949 | −2.4009 | −0.4908 | 0.0415 | 2.0300 |

Table 2. Properties of $Al_xTi_yNi_z$ clusters with $x + y + z = 6$. Numbers in the first column indicate the number of Al, Ti and Ni atoms in each cluster. The average interatomic distances (AID), binding energy per atom ($E_b$), excess energy ($E_{\mathrm{exc}}$), second order difference energy ($\Delta_E$), and HOMO–LUMO energy gaps ($E_{\mathrm{gap}}$) are presented in the table. The unit of AID is Angstrom (Å) and $E_b$, $E_{\mathrm{exc}}$, $\Delta_E$ and $E_{\mathrm{gap}}$ are in the unit of electronvolt ($eV$).

| Alloy | $N$–$N$ pairs | $\sigma$ | IP | EA | $\eta$ | $\xi$ |
|-------|---------------|----------|-----|-----|--------|-------|
| 006 | 0 0 9 0 0 0 | 1 | 5.0352 | 1.9009 | 3.1343 | 3.4680 |
| 060 | 0 12 0 0 0 0 | 1 | 4.9202 | 1.6542 | 3.2660 | 3.2872 |
| 600 | 12 0 0 0 0 0 | 1 | 6.5081 | 2.2685 | 4.2395 | 4.3883 |
| 015 | 0 0 5 0 0 5 | 0 | 3.6124 | 2.8348 | 0.7776 | 3.2236 |
| 024 | 0 1 4 0 0 8 | −0.2308 | 5.7500 | 1.0726 | 4.6773 | 3.4113 |
| 033 | 0 3 0 0 0 8 | −0.4545 | 4.9424 | 1.2758 | 3.6667 | 3.1091 |
| 042 | 0 6 0 0 0 6 | 0 | 4.6871 | 1.3275 | 3.3595 | 3.0073 |
| 051 | 0 9 0 0 0 3 | 0.5 | 4.3044 | 1.4923 | 2.8121 | 2.8983 |
| 105 | 0 0 7 5 0 0 | 0.1667 | 8.8293 | −1.4159 | 10.2453 | 3.7067 |
| 204 | 1 0 3 8 0 0 | −0.3333 | 4.1142 | 1.7507 | 2.3635 | 2.9325 |
| 303 | 3 0 0 7 0 0 | −0.4 | 7.4341 | −0.0486 | 7.4826 | 3.6927 |
| 402 | 4 0 0 7 0 0 | −0.2728 | 5.8154 | 1.4024 | 4.4130 | 3.6089 |
| 501 | 5 0 0 5 0 0 | 0 | 5.9898 | 1.8785 | 4.1113 | 3.9341 |
| 150 | 0 9 0 0 3 0 | 0.5 | 5.9271 | 0.5799 | 5.3473 | 3.2535 |
| 240 | 0 4 0 0 8 0 | −0.3333 | 5.9192 | 1.2855 | 4.6336 | 3.6023 |
| 330 | 0 3 0 0 8 0 | −0.45 | 6.2936 | 1.3395 | 4.9541 | 3.8166 |
| 420 | 3 1 0 0 8 0 | −0.3333 | 5.4543 | 1.9843 | 3.4700 | 3.7193 |
| 510 | 6 0 0 0 4 0 | 0.2 | 6.1389 | 1.4629 | 4.6760 | 3.8009 |
| 114 | 0 0 4 4 1 4 | −0.3846 | 6.3526 | 0.9253 | 5.4273 | 3.6390 |
| 123 | 0 1 0 2 2 6 | −0.8181 | 5.4550 | 1.0456 | 4.4094 | 3.2503 |
| 132 | 0 3 0 2 3 4 | −0.5 | 5.2388 | 0.8214 | 4.4174 | 3.0301 |
| 141 | 0 5 0 1 3 2 | −0.9090 | 5.5875 | 0.8501 | 4.7374 | 3.2188 |
| 213 | 1 0 0 5 2 3 | −0.8181 | 5.5041 | 1.0529 | 4.4512 | 3.2785 |
| 222 | 0 1 0 3 4 4 | −0.8333 | 5.2723 | 1.6985 | 3.5739 | 3.4854 |
| 231 | 0 3 0 1 6 2 | −0.5 | 5.6497 | 1.6794 | 3.9703 | 3.6645 |
| 312 | 1 0 0 4 3 2 | −0.8 | 5.9228 | 1.5396 | 4.3832 | 3.7312 |
| 321 | 1 1 0 2 6 2 | −0.6667 | 6.0685 | 1.4751 | 4.5934 | 3.7718 |
| 411 | 2 0 0 3 4 1 | −0.6 | 6.2539 | 1.7434 | 4.5104 | 3.9987 |

Table 3. Properties of $Al_xTi_yNi_z$ clusters with $x + y + z = 6$. Numbers in the first column indicate the number of Al, Ti and Ni atoms in each cluster. The number of nearest neighbor pairs (in sequence of Al–Al, Ti–Ti, Ni–Ni, Al–Ni, Al–Ti and Ti–Ni), chemical order ($\sigma$), ionization potential (IP), electron affinity (EA), global hardness ($\eta$) and Mulliken electronegativity ($\xi$) are presented in the table. There is no unit for $\sigma$ and the unit of IP, EA , $\eta$ and $\xi$ is electronvolt ($eV$).

| Alloy | $M$ | $E_T$ | $E_T + ZPE$ | $\langle\alpha\rangle$ | $\langle\alpha\rangle/N$ |
|-------|-----|-------|-------------|------------------------|--------------------------|
| 006 | 1 | −246241.342 | −246241.188 | 188.256 | 31.376 |
| 060 | 1 | −138677.335 | −138677.160 | 297.875 | 49.646 |
| 600 | 1 | −39583.427 | −39583.280 | 281.389 | 46.898 |
| 015 | 1 | −228316.983 | −228316.832 | 184.386 | 30.731 |
| 024 | 1 | −210391.496 | −210391.347 | 222.542 | 37.090 |
| 033 | 1 | −192463.490 | −192463.336 | 233.030 | 38.838 |
| 042 | 1 | −174535.035 | −174534.866 | 250.744 | 41.791 |
| 051 | 1 | −156606.180 | −156606.014 | 273.329 | 45.555 |
| 105 | 2 | −211804.322 | −211804.182 | 225.358 | 37.560 |
| 204 | 1 | −177360.405 | −177360.224 | 223.446 | 37.241 |
| 303 | 2 | −142918.303 | −142918.131 | 250.248 | 41.708 |
| 402 | 1 | −108473.823 | −108473.655 | 247.223 | 41.204 |
| 501 | 2 | −74028.739 | −74028.594 | 284.405 | 47.401 |
| 150 | 2 | −122163.037 | −122162.889 | 309.165 | 51.528 |
| 240 | 1 | −105647.403 | −105647.214 | 277.511 | 46.252 |
| 330 | 2 | −89131.989 | −89131.8391 | 302.352 | 50.392 |
| 420 | 1 | −72615.405 | −72615.2587 | 308.101 | 51.350 |
| 510 | 2 | −56099.640 | −56099.497 | 303.740 | 50.623 |
| 114 | 2 | −193877.180 | −193877.013 | 221.678 | 36.946 |
| 123 | 2 | −175949.085 | −175948.931 | 247.314 | 41.219 |
| 132 | 2 | −158020.397 | −158020.237 | 271.550 | 45.258 |
| 141 | 2 | −140091.496 | −140091.331 | 270.691 | 45.115 |
| 213 | 1 | −159433.846 | −159433.676 | 237.198 | 39.533 |
| 222 | 1 | −141505.107 | −141504.941 | 257.557 | 42.926 |
| 231 | 1 | −123576.476 | −123576.291 | 267.729 | 44.622 |
| 312 | 2 | −124990.021 | −124989.843 | 253.264 | 42.211 |
| 321 | 2 | −107060.933 | −107060.769 | 287.704 | 47.951 |
| 411 | 1 | −90545.009 | −90544.846 | 282.451 | 47.075 |

Table 4. Properties of $Al_xTi_yNi_z$ clusters with $x + y + z = 6$. Numbers in the first column indicate the number of Al, Ti and Ni atoms in each cluster. Followed by multiplicity ($M$), total energy ($E_T$), summation of total energy and zero point energy ($E_T + ZPE$), static mean polarizability ($\langle\alpha\rangle$) and mean polarizability ($\langle\alpha\rangle/N$). The unit of the $E_T$ and $E_T + ZPE$ is electronvolt ($eV$), whereas the unit of $\langle\alpha\rangle$ and $\langle\alpha\rangle/N$ is atomic unit ($a.u.$).

**Definitions of ionization potential (IP) and electron affinity (EA) in the context of the $Al_xTi_yNi_z$ clusters**

Ionization potential (IP) is defined as the total energy difference between the electronic ground structure of the cationic and the neutral cluster structures,

$$\text{IP} = E_{Al_xTi_yNi_z} - E_{Al_xTi_yNi_z^+} \, .$$

Electron affinity (EA) is defined as the total energy difference between the neutral and anionic cluster structures,

$$\text{EA} = E_{Al_xTi_yNi_z^-} - E_{Al_xTi_yNi_z}.$$

**Definitions of global hardness ($\eta$) and Mulliken electronegativity ($\xi$)**

Global hardness ($\eta$) and Mulliken electronegativity ($\xi$) are two quantitative parameters measuring the chemical reactivity of a given specific cluster in a charge transfer process. They are defined in terms of IP and EA, as per

$$\eta = \text{IP} - \text{EA} \, ,$$

$$\xi = \frac{1}{2}(\text{IP} + \text{EA}) \, .$$